\documentclass[a4paper,11pt]{article}
\usepackage{pos}
\title{Heavy flavor production in the Parton-Hadron-String Dynamics (PHSD)}
\author*[a]{Taesoo Song}
\author[b,c]{Joerg Aichelin}
\author[a,d,e]{Elena Bratkovskaya}

\affiliation[a]{GSI Helmholtzzentrum f\"{u}r Schwerionenforschung GmbH,\\
  Planckstrasse 1, 64291 Darmstadt, Germany}

\affiliation[b]{SUBATECH UMR 6457 (IMT Atlantique,  Universit\'{e} de Nantes, IN2P3/CNRS),\\
4 Rue Alfred Kastler, 44307 Nantes, France}

\affiliation[c]{Frankfurt Institute for Advanced Studies,\\
Ruth-Moufang-Strasse 1, 60438 Frankfurt am Main, Germany}

\affiliation[d]{Institut f\"ur Theoretische Physik, Johann Wolfgang Goethe-Universit\"at,\\
Max-von-Laue-Str.\ 1, D-60438 Frankfurt am Main, Germany}

\affiliation[e]{Helmholtz Research Academy Hesse for FAIR (HFHF),\\
GSI Helmholtz Center for Heavy Ion Physics, Campus Frankfurt, 60438 Frankfurt, Germany}

\emailAdd{T.Song@gsi.de}
\emailAdd{aichelin@subatech.in2p3.fr}
\emailAdd{E.Bratkovskaya@gsi.de}

\abstract{Relativistic heavy-ion collisions produce a hot and dense nuclear matter, through which one can study the phase diagram of QCD.
Open and hidden heavy flavors are promising probes to search for the properties of the 
hot and dense nuclear matter under extreme conditions.
We present how the production and interactions of open and hidden heavy flavors in heavy-ion collisions are realized in the Parton-Hadron-String Dynamics, which is
a non-equilibrium microscopic transport approach for the description of the dynamics of strongly interacting hadronic and partonic matter. 
}

\FullConference{20th International Conference on B-Physics at Frontier Machines (Beauty2023)\\
 3-7 July, 2023\\
Clermont-Ferrand, France\\}

\begin{document}
\maketitle

\section{Introduction}
Relativistic heavy-ion collisions produce nuclear matter in extreme conditions.
%At high energy collisions two nuclei pass through each other, and a large number of (anti)particles are produced.
%So the produced matter is located at high temperature and low baryon chemical potential in the QCD phase diagram.
%On the other hand, low energy collisions produce a nuclear matter at low temperature and high baryon chemical potential, because two colliding nuclei stop and are compressed.
%In this case only a small number of (anti)particles are produced and the colliding nuclei remain in the middle.
The properties of the matter produced in heavy-ion collisions can be probed by bulk particles such as pions and kaons, electromagnetic particles and hard particles like jets and heavy flavors.
Heavy flavors have the following advantages:
First, their production requires a large energy-momentum transfer, i.e. a hard scattering.
Thus the production can be described in pQCD and is model-independent.
Second, since heavy flavors are produced early in heavy-ion collisions, they potentially contain the information of the produced matter at the initial stage of heavy-ion collisions.
In fact heavy flavors are not fully thermalized in heavy-ion collisions due to their large mass and some memory of the initial stage survives.

In experiment it is observed that  single electrons from $B$ meson decays are suppressed at large transverse momentum~\cite{ALICE:2022iba} and the excited states of bottomonium are strongly suppressed - compared to the ground state \cite{CMS:2011all} - in heavy-ion collisions at LHC. 
The former attributes to the interaction of heavy quarks with partons in the QGP, which leads to an energy loss of fast heavy quarks, and the latter to the dissolution of excited states in a QGP at high temperature.
Therefore, both reveal the properties of matter in extreme conditions produced in relativistic heavy-ion collisions.

\section{Parton-hadron-string dynamics (PHSD)}

For more precise quantitative studies of heavy flavor production and dynamics it is necessary to describe the space-time evolution of the matter produced in heavy-ion collisions.
The Parton-Hadron-String Dynamics (PHSD) is a non-equilibrium microscopic transport approach for the description of the dynamics of strongly interacting hadronic and partonic matter~\cite{Cassing:2009vt}.
The dynamics is based on the solution of generalized off-shell transport equations derived from Kadanoff-Baym many-body theory which is beyond the semi-classical BUU transport.
 
Heavy-ion collisions in the PHSD start by nucleon-nucleon scattering.
When the collision energy is relatively low, it leads to elastic scattering or the excitation of nucleon(s) or at most the production of a couple of particles.
As the collision energy increases, however, more and more particles are produced and the LUND string model is useful to describe the multiparticle production. The number of strings grows with increasing bombarding energies. 
%In heavy-ion collisions many strings are produced and each string has a larger energy with increasing collision energy.
In PHSD, if the local energy density is above the critical value for a phase transition (according to the lQCD), strings melt into quarks and antiquarks and form a QGP.
Otherwise, strings fragment into hadrons, which is called string fragmentation.

The equation-of-state (EoS) of the QGP at small baryon chemical potential is available from lattice QCD calculations.
However, the EoS is a macroscopic property of the QGP and  does not provide a microscopic picture as well as information on the degrees-of-freedom of the QGP.
Therefore, PHSD adopts the dynamical quasi-particle model (DQPM) where (anti)quark and gluon are expressed by spectral functions whose pole masses and spectral widths depend on temperataure and baryon chemical potential~\cite{Moreau:2019vhw}.
The  pole mass and spectral widths follow those from the hard thermal loop calculations but the strong coupling is extracted from a lattice EoS in order to access the non-perturbative region close to $T_c$.
As a result, the DQPM satisfies lattice EoS and additionally provides a microscopic view of the QGP.
Furthermore, the micriscopic picture enables us to calculate transport coefficients of the QGP which are consistent with lattice QCD results in thermodynamic equilibrium \cite{Soloveva:2019xph}.

When the QGP expands with time and the temperature lowers down to $T_c$, partons hadronize, while conserving energy-momentum and all quantum numbers.
Since partons are off-shell, the hadronized mesons and (anti)baryons are also off-shell by energy-conservation.
The scattering of off-shell hadrons is based on many-body theory (G-matrix), chiral models or experimental data on scattering cross sections. We recall that the 
PHSD provides a good description of many bulk observables such as rapidity and transverse momentum distributions and flow coefficients of hadrons from SchwerIonen Synchrotron (SIS) to Large Hadron Collider (LHC) energies~\cite{Linnyk:2015rco}.

\section{Open heavy flavor production}

In PHSD the production point and initial momentum of a heavy quark pair are provided  by the Glauber model and the PYTHIA event generator~\cite{Sjostrand:2006za}.
Since the PYTHIA event generator is based on leading-order pQCD calculations, though intial and final showers are included, the momentum and rapidity of heavy quarks from  PYTHIA are rescaled such that their distributions are similar to those from the fixed-order next-to-leading logarithm (FONLL)
calculations~\cite{Cacciari:2005rk}.
In $pp$ collisions the produced heavy quark hadronizes through heavy quark fragmentation by using the Peterson's function~\cite{Peterson:1982ak} for transverse momentum and its chemistry is decided based on the experimental data which are generally energy-independent.
 
The parton distribution function (PDF) in a nucleus is modified from that in a single nucleon, which is called (anti)shadowing effect.
This modification affects heavy flavor production, because the main process are $g+g\rightarrow Q+\bar{Q}$ or $q+\bar{q}\rightarrow Q+\bar{Q}$.
This (anti)shadowing effect is realized in PHSD using the EPS09 package~\cite{Eskola:2009uj}.

Heavy quarks interact in the QGP through elastic scattering with off-shell massive partons.
The scattering cross sections are calculated by leading-order Feynman diagrams based on the DQPM, where the strong coupling depends on temperature and baryon chemical potential, while the off-shell mass of the exchanged parton plays the role of a regulator.
The resulting spatial diffusion coefficient is consistent with that from lattice QCD calculations~\cite{Berrehrah:2014kba}.

As the local energy density approaches 0.75 ${\rm GeV/fm^3}$ during the expansion of the  matter, a heavy (anti)quark tries coalesence with light partons close both in coordinate and in momentum spaces~\cite{Song:2015sfa}.
If a heavy (anti)quark fails coalesence until the energy density drops below than 0.4 ${\rm GeV/fm^3}$, it is forced to hadronize by fragmentation as in $pp$ collisions.
For an energetic heavy quark, it can hardly find a neighboring light quark in momentum space, though it is surrounded by them in coordinate space.
Therefore the coalescence probability is large at low momentum and decreases with increasing heavy quark momentum.
The probability also decreases with increasing the centrality of heavy-ion collisions, and the hadronization process will dominantly be fragmentation in extremely peripheral collisions.
The hadronized $D$ mesons interact with light mesons and baryons with  scattering cross sections calculated in an effective chiral lagrangian approach with heavy-quark spin symmetry which is  unitarized~\cite{Song:2015sfa}.

\begin{figure}[h]
\centerline{
\includegraphics[width=8.3 cm]{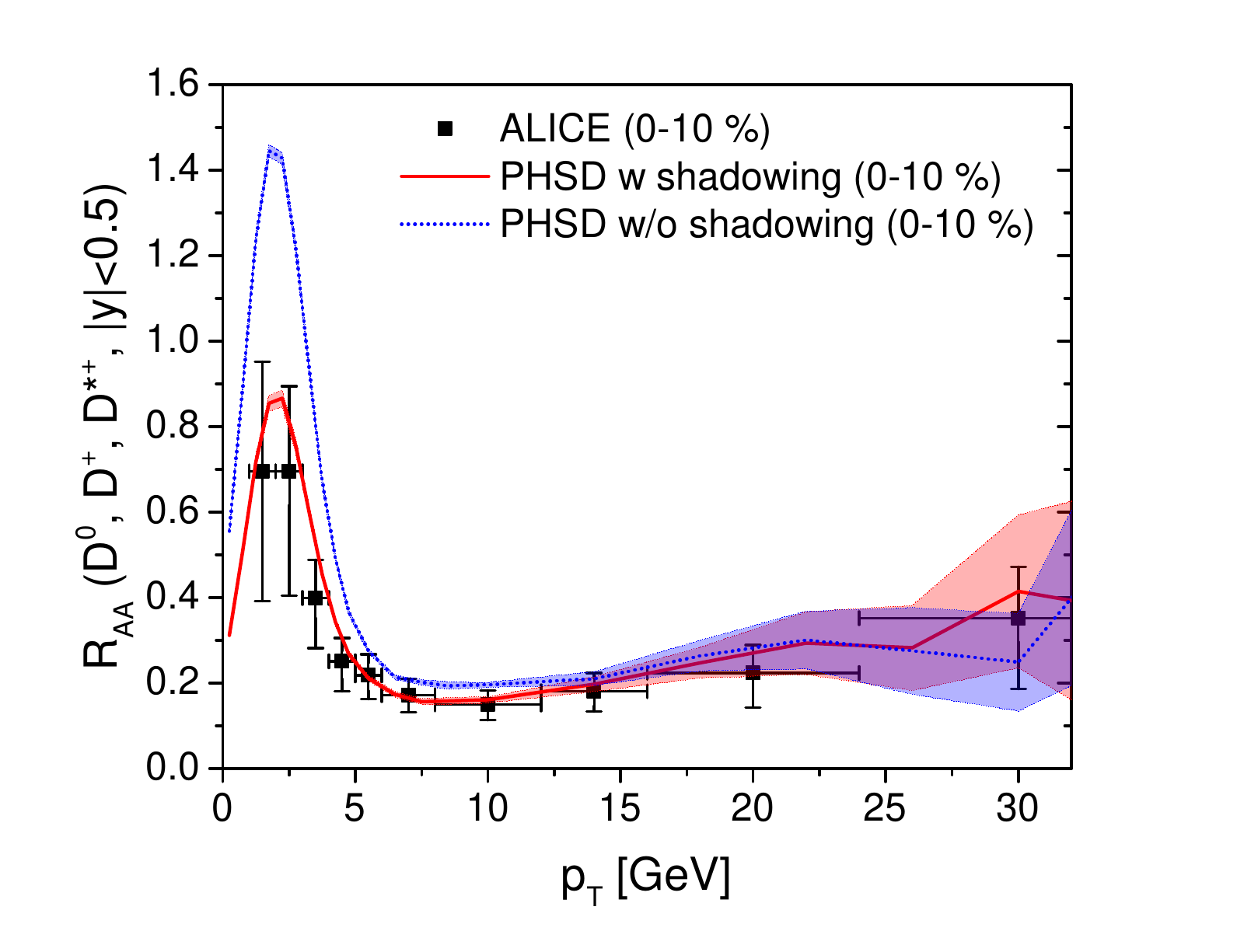}
\includegraphics[width=8.3 cm]{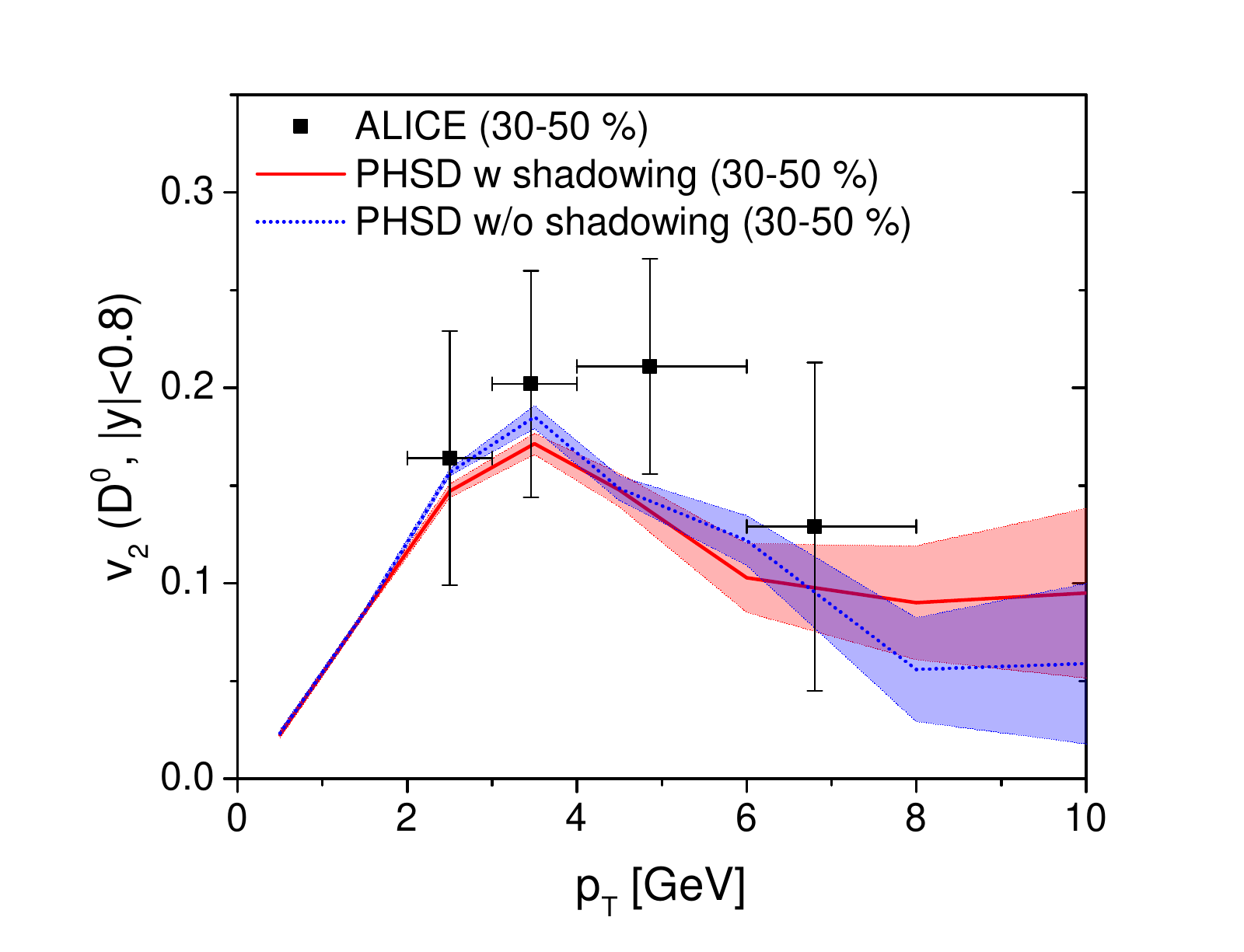}}
\caption{(Left) $R_{\rm AA}$ and (right) $v_2$ of $D$ mesons respectively in 0-10 \% and 30-50 \% central Pb+Pb collisions at $\sqrt{s_{\rm NN}}=$ 2.76 TeV from the PHSD~\cite{Song:2015ykw}, compared to experimental data from the ALICE Collaboration~\cite{ALICE:2015vxz}.}
\label{raa-D}
\end{figure}

Fig.~\ref{raa-D} shows the PHSD results for the nuclear modification factor $R_{\rm AA}$ and elliptic flow $v_2$ of $D$ mesons, respectively, in 0-10 \% and 30-50 \% central Pb+Pb collisions at $\sqrt{s_{\rm NN}}=$ 2.76 TeV from the PHSD~\cite{Song:2015ykw} in comparison to the experimental data from the ALICE Collaboration~\cite{ALICE:2015vxz}.
One can see that the (anti)shadowing effect is necessaray to explain $R_{\rm AA}$ in central collisions but the influence on $v_2$ is insignificant.

\section{Hidden heavy flavor production}

\begin{figure}[h]
\centerline{
\includegraphics[width=8.3 cm]{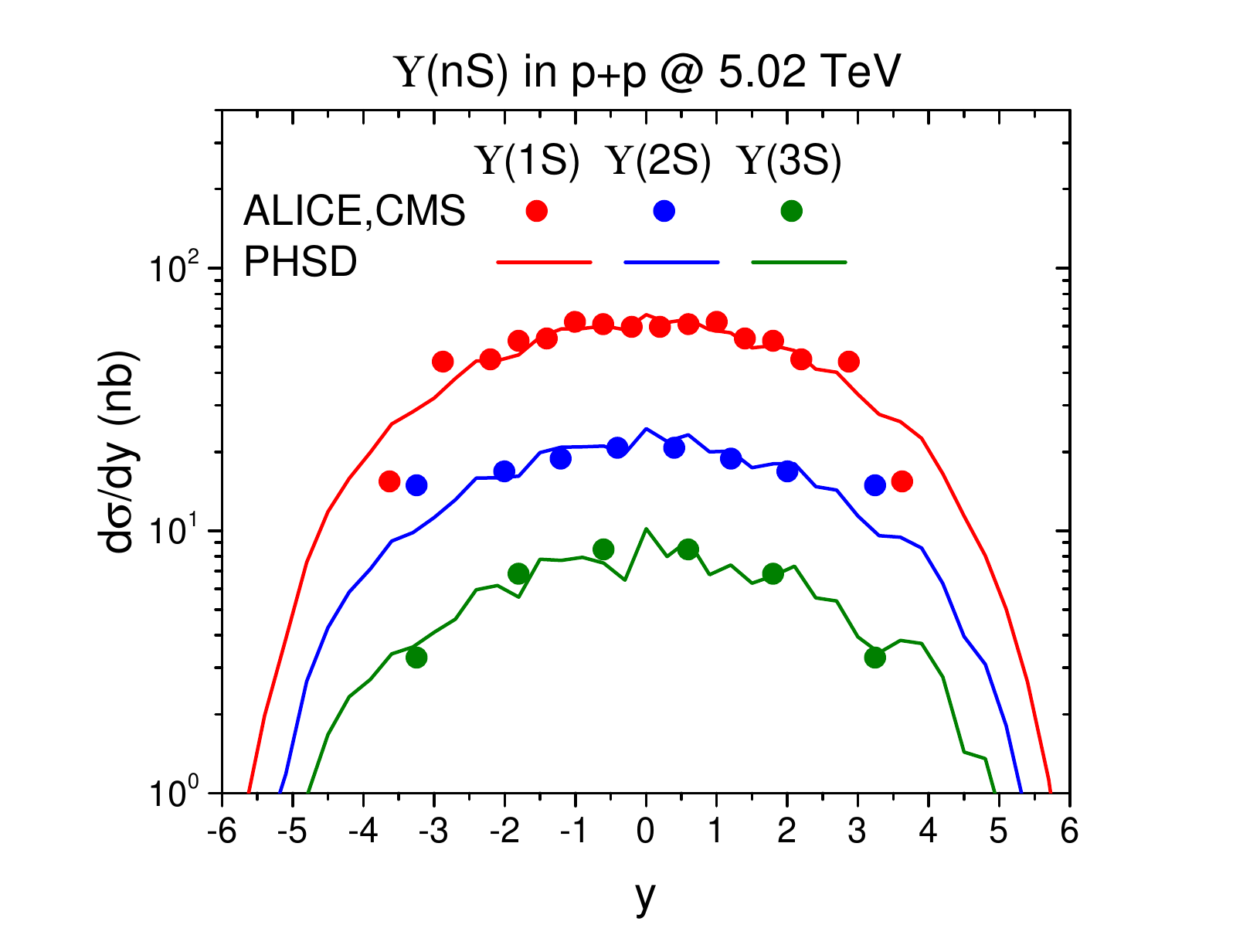}
\includegraphics[width=8.3 cm]{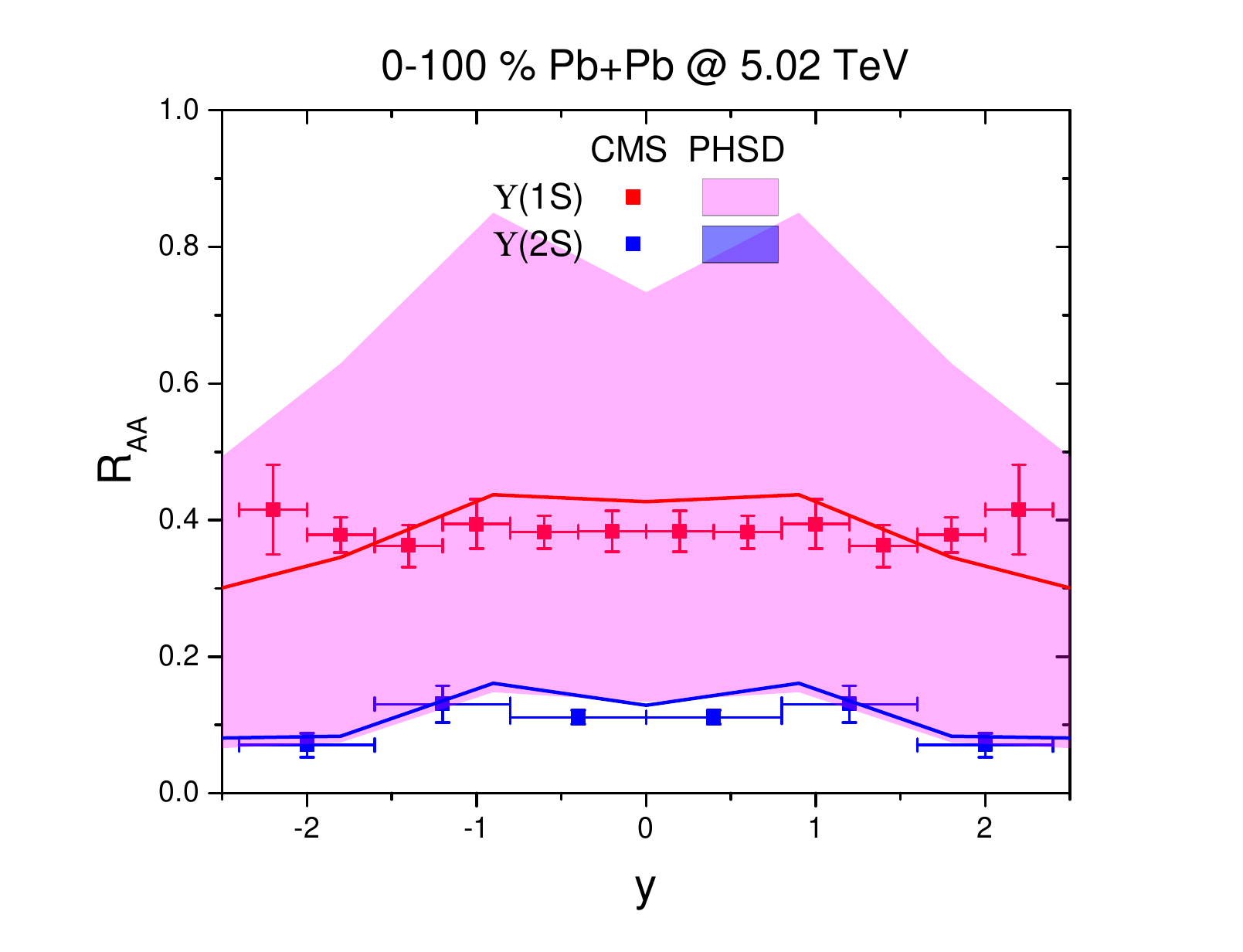}}
\caption{(Left) The PHSD results for the rapidity distribution of $\Upsilon$ (nS) in $pp$ collisions at $\sqrt{s_{\rm NN}}=$ 5.02 TeV and (right) $R_{\rm AA}$ of $\Upsilon$ (1S) and $\Upsilon$(2S) for the free energy potential in 0-80 \% central Pb+Pb collisions at the same energy, compared with experimental data from the CMS and ALICE Collaborations~\cite{CMS:2018zza,ALICE:2021qlw}.}
\label{raa}
\end{figure}

Quarkonium production in $pp$ collisions takes two steps: First, a heavy quark pair is produced by a hard scattering and then the produced pair forms a bound state.
The former is a hard process where pQCD is applicable, while the latter is a soft process which must depend on a suitable model.
In PHSD the former is realized by the PYTHIA event generator and the latter is based on the Wigner projection where the only parameter is the quarkonium radius.
It successfully describes charmonium and bottomonium production in $pp$ collisions~\cite{Song:2017phm,Song:2023zma} as shown in the left panel of Fig.~\ref{raa}.

In heavy-ion collisions there are a couple of differences from in $pp$ collisions.
First, the quarkonium radius depends on temperature because the heavy quark potential changes in the QGP.
We take the free energy of the heavy quark system from lattice QCD calculations for the heavy quark potential~\cite{Lee:2013dca} and solve the Schr\"{o}dinger equation to obtain the eigenvalue and eigenfunction of the each state of quarkonium.
Second, a heavy quark and a heavy antiquark - from two different initial pairs - may form a bound state in heavy-ion collisions, which is called the off-diagonal recombination. 

For the description of quarkonium production in heavy-ion collisions, 
we  use Remler's formalism, where the Wigner projection is carried out first, when local temperature of the QGP lowers down to the dissociation temperataure of each quarkonium state, and it is updated whenever a heavy quark or heavy antiquark interacts in the QGP~\cite{Song:2023zma}.

The right panel of Fig.~\ref{raa} shows $R_{\rm AA}$ of $\Upsilon$(1S) and $\Upsilon$(2S) in 0-80 \% central Pb+Pb collisions at $\sqrt{s_{\rm NN}}=$ 5.02 TeV.
The upper limit of the magenta band indicates the initial production of $\Upsilon$(1S)  at its dissociation temperature which is about 3.3 $T_c$ for the free energy heavy quark potential.
On the other hand, the lower limit displays the final $\Upsilon$(1S) at the end of the QGP, which shows that heavy quark interactions in the QGP suppresses $\Upsilon$(1S).
One can see that the experimental data from the CMS Collaboration~\cite{CMS:2018zza} are between the upper and lower limits.
Assuming that only 10 \% of heavy (anti)quark scattering in the QGP affects bottomonium production and dissociation, one gets the red solid line which is consistent with the CMS data. 
%It is reasonable in sense that if heavy quark and heavy antiquark (form $\Upsilon$ (1S)), its scattering cross section must be much smaller than twice the heavy quark scattering cross section, because the $\Upsilon$ (1S) is a color-singlet with a small size.
It is reasonable to conclude that a heavy quark and heavy antiquark (form $\Upsilon$(1S)) scattering cross section must be much smaller than twice the heavy quark scattering cross section, because the $\Upsilon$(1S) is a color-singlet with a small size.
On the other hand, $\Upsilon$(2S) has a very narrow band, because its dissociation temperature is close to $T_c$.

\section{Summary}
The PHSD describes a hadronic and partonic matter produced in heavy-ion collisions in terms of strongly interacting off-shell particles which reproduce lattice and expermental data.
Open heavy flavor production in heavy-ion collisions is realized in PHSD by help of the PYTHIA event generator for the initial production, the EPS09 package for (anti)shadowing effects,  the DQPM for interaction cross sections in QGP, the coalescence and fragmentation for hadronization and the effective chiral lagrangian with unitarizataion for hadronic scattering.
According to Remler's formalism quarkonium production is closely related to the interaction of open heavy flavors in QGP, and we have found that the scattering cross section of two open heavy flavors reduces to about 10\% for hidden heavy flavor in heavy-ion collisions. 

\section*{Acknowledgements}
We acknowledge support by the Deutsche Forschungsgemeinschaft (DFG, German Research Foundation) through the grant CRC-TR 211 'Strong-interaction matter under extreme conditions' - Project number 315477589 - TRR 211. 
This work is supported by the European Union’s Horizon 2020 research and innovation program under grant agreement No 824093 (STRONG-2020).
The computational resources have been provided by the LOEWE-Center for Scientific Computing and the "Green Cube" at GSI, Darmstadt and by the Center for Scientific Computing (CSC) of the Goethe University.

\end{document}